\documentclass[10pt]{article}

\usepackage{amssymb}
\usepackage{graphicx}
\usepackage[numbers,compress]{natbib}
	\author{Ion I. Cot\u aescu\thanks{Corresponding author E-mail:~~i.cotaescu@e-uvt.ro}\\
	{\it West University of Timi\c soara,} \\{\it V. Parvan Ave. 4,
		RO-300223 Timi\c soara}}

\begin{document}
\title{Evaporating black holes in de Sitter expanding universe}

\maketitle

\begin{abstract}
Models of evaporating black holes are constructed using the new  solutions of  Einstein's equations with perfect fluid in space-times with FLRW asymptotic behaviour  derived recently   [I. I. Cotaescu, Eur. Phys. J. C  (2022) 82:86].  The dynamics of these models is  exclusively due to the interplay between black holes and their environments, without resorting to additional matter sources or thermodynamic considerations.  During evaporation the black hole mass dissipates into a cloud of dust which  replaces the black hole while the background expands tending to the asymptotic one.

Pacs: 04.70.Bw
\end{abstract}


\section{Introduction}

 The Friedmann- Lema\^ itre-Robertson-Walker (FLRW) universes may be populated  either with static black holes, which are vacuum solutions of Einstein's equations \cite{BH}, or with dynamical particles defined as isotropic solutions of Einstein's equations with perfect fluid laying out a FLRW asymptotic behaviour (see for instance Ref. \cite{TB1}). When one studies a static black hole in a given background one must consider the black hole and its FLRW background as generated by different gravitational sources. For example, in the case of the Schwarzschild-de Sitter black hole (having the Kottler metric \cite{Kot}) the de Sitter background is generated by an independent  cosmological constant. 
 
Many authors studied models  describing coherently the evolution of dynamical particles in FLRW backgrounds that can behave as black holes on some space-times domains.  The first such model  was proposed by McVittie long time ago \cite{MV} and then studied by many authors \cite{MV1,MV2,MV3}. Recently the McVittie geometry was generalized to new solutions of the Einstein equations with perfect fluid behaving asymptotically as FLRW space-times with curved space sections \cite{MV4}. The general feature of these models is that their gravitational source is the fluid pressure which is singular on the Schwarzschild sphere while the density remains just that of the asymptotic FLRW space-time.

Alternatively,  we proposed recently  a new type of dynamical particles which are exact solutions of Einstein's equations with perfect fluid preserving the fluid pressure of the asymptotic FLRW space-times  but this time with flat space sections  \cite{Cot}. We have shown that these dynamical particles behave as non-rotating black holes in the physical space domain bordered by the black hole and cosmological dynamical horizons. Moreover, these dynamical particles give rise to photon spheres and black hole shadows as in the case of the static Schwarzschild black holes.  For this reason we say here that these are  Schwarzschild-type dynamical particles or black holes. Subsequently we generalized these models obtaining new solutions of Einstein's equations with perfect fluid depending on a time-dependent black hole mass and a real valued parameter $\kappa$ giving more flexibility to these models, we call simply the $\kappa$-models \cite{Cot1}. The solutions we studied previously \cite{Cot} are  particular cases with $\kappa=0$. 

In the present paper we would like to initiate the study of the physical meaning of the dynamical quantities of  the $\kappa$-models. We show that these models describe systems constituted by a Schwarzschild-type black holes surrounded by a cloud of dust, hosted by the perfect fluid determining the FLRW asymptotic behaviour of the background, which we call here the FLRW fluid. Moreover, we show that when this background  is expanding then the black hole evaporates dissipating its mass into dust but without affecting the FLRW fluid.  Remarkably, in these models the Hubble function of the FLRW fluid is proportional through $\kappa$ to the total mass of the dynamical system black hole-dust.  This property offer us the opportunity of studying here the models in which this mass is conserved and consequently  the asymptotic FLRW manifold  is just the de Sitter one,  having a constant Hubble function given now by our parameter $\kappa$ instead of a hypothetical cosmological constant. Studying these models we may understand how  these black holes evaporate naturally because of their geometries without involving supplemental mechanism as, for example,  the Hawking radiation.  
  
We start our study in the next section presenting the framework and the principal notations concerning the form of Einstein's equations with perfect fluid in physical frames showing that their solutions have a FLRW asymptotic behaviour. The next section is devoted to our $\kappa$-models and their components that are briefly analyzed. The principal new result of this section is the relation between the total mass of the system black hole-dust and the Hubble function of the asymptotic FLRW space-time. Based on this relation we define in section 4 the models with conserved total mass whose asymptotic space-times are just the de Sitter ones whose perfect fluid is assumed to be dark energy. We study here the black hole evaporation and the dynamics of the whole geometry laying out black hole, cosmological and asymptotic horizons, the last ones being  de Sitter event horizons. It is interesting that the black hole and cosmological horizons can form either usual C-curves or open curves with parallel asymptotes open to $t\to-\infty $.  In both these cases the ordinary masses of the black holes evaporates completely dissipating into dust.   Some concluding remarks are presented in the last section. 

An integral and the method of solving cubic equations are given  in two Appendices.

 We use the Planck units with $\hbar=c=G=1$.

\section{Dynamical particles in physical frames}

The static or dynamic non-rotating black holes are studied mainly in space-times with spherical symmetry and spatially flat sections. In these manifolds it is convenient to use the physical frames $\{t, {\bf x}\}$  with  Painlev\' e-Gullstrand  coordinates  \cite{Pan,Gul},  $x^{\mu}$ ($\alpha,\mu,\nu,...=0,1,2,3$) formed by the {\em cosmic time}, $x^0=t$, and physical Cartesian space coordinates,  ${\bf x}=(x^1,x^2,x^3)$, associated to the spherical ones $(r,\theta,\phi)$. For example, in the physical frame of the Schwarzschild-de Sitter black hole, the Kottler (\cite{Kot}) line element takes the form 
\begin{equation}
ds^2=f(r)dt^2+2\sqrt{1-f(r)}\,dtdr-dr^2 -r^2d\Omega^2\,,\label{ss}
	\end{equation}	
where 
\begin{equation}\label{frSdS}
	f(r)=1-\frac{2m}{r}-\omega_{H}^2 r^2\,,
\end{equation} 
depends on the black hole mass $m$ and Hubble de Sitter constant frequency $\omega_{H}$.  Eq. (\ref{ss}) suggests us the substitution $f(r)=1-h(r)^2$ helping us to simplify the off-diagonal components of the metric tensor. Therefore, in what follows  we consider that the line elements in physical frames of the dynamical manifolds have the general form  
\begin{eqnarray}
	ds^2&=& g_{\mu\nu}(x)dx^{\mu}dx^{\nu}=dt^2 -\left[dr-h (t,r)dt\right]^2-r^2 d\Omega^2 \nonumber\\
	&=&\left[1-h (t,r)^2\right]dt^2+2 h(t,r)dr dt -dr^2-r^2 d\Omega^2\,,\label{fam}
\end{eqnarray} 
depending on the smooth functions $h$ and $d\Omega^2=d\theta^2+\sin^2\theta\, d\phi^2$. In these frames we have  $g^{00}=1$, $g_{0r}=g^{0r}=h$ and 
\begin{equation}\label{gg}
\sqrt{g}=\sqrt{-\det(g_{\mu\nu})}=r^2\sin\theta\,,
\end{equation}
which may simplify some calculations. 

The functions $h$ have to be derived solving the Einstein equations in the sense of selecting  the  manifolds ${\frak M}$ whose $h$-functions give isotropic Einstein tensors that can be expressed as  
\begin{equation}\label{Ein}
	G^{\mu}_{\,\nu}=\Lambda\delta^{\mu}_{\nu}+ 8\pi\left[ (\rho+p)U^{\mu}U_{\nu}-p\delta^{\mu}_{\nu}\right]\,,
\end{equation}
in terms of a possible cosmological constant $\Lambda$ and the density $\rho$ (of matter or energy) and pressure $p$ of a perfect fluid  moving with the four-velocity $U^{\mu}$ with respect to the physical  frame under consideration. In a proper co-moving frame where the four-velocity has the components 
\begin{eqnarray}
	U_{\mu}&=&\left(\frac{1}{\sqrt{g^{00}}},0,0,0\right)=(1,0,0,0)\,, \\ 
	U^{\mu}&=&g^{\mu 0}U_0=\frac{g^{\mu 0}}{\sqrt{g^{00}}} ~~\Rightarrow~~U^{\mu}=(1,h,0,0)\,,
\end{eqnarray}
Eqs. (\ref{Ein}) are solved by an isotropic Einstein tensor  whose non-vanishing components satisfy
\begin{eqnarray}
	&&G^r_r=G^{\theta}_{\theta}=G^{\phi}_{\phi}\equiv G\,,\label{iso}\\
	G^0_r=0 &\Rightarrow&G^r_0=\frac{g^{0r}}{g^{00}}\left(G^0_0-G\right)\,,\label{cond}
\end{eqnarray}
defining the gravitational sources  in the co-moving frame as
\begin{eqnarray}
	G^0_{0} &=&\Lambda + 8\pi \,\rho\,, \label{E10}\\
	G&=&\Lambda - 8\pi \, p \,.\label{E20}
\end{eqnarray}
Hereby we may  identify the density  $\rho$ and pressure $p$ of the perfect fluid of ${\frak M}$  resorting eventually to additional suppositions for separating a contribution of the cosmological constant $\Lambda$.    

The asymptotic space-time of  ${\frak M}$, for $r\to\infty$,  is a FLRW manifold, ${\frak M}(a)$, of the scale factor $a(t)$ whose Hubble function is defined by the asymptotic condition 
\begin{equation}
	\frac{1}{a(t)}\frac{da(t)}{dt}\equiv\frac{\dot a(t)}{a(t)}=\lim_{r\to\infty}\frac{h(t,r)}{r}\,.
\end{equation}
The metric tensors of  ${\frak M}(a)$, give the line elements in physical frames of the form,
\begin{equation}\label{s2}
	ds^2=\left(1-\frac{\dot a^2}{a^2}\, r^2\right)dt^2+2\frac{\dot a}{a}\, r\, dr\, dt -dr^2-r^2d\Omega^2\,,
\end{equation}
 and satisfy the Friedmann equations,
\begin{eqnarray}
	G^0_{0}(a) &=&3\left(\frac{\dot a}{a}\right)^2=\Lambda + 8\pi \,\rho_a\,, \label{E1}\\
	G(a)&=&3\left(\frac{\dot a}{a}\right)^2 +2\frac{d}{dt}\left(\frac{\dot a}{a}\right) =\Lambda - 8\pi \, p_a \,,\label{E2}
\end{eqnarray}
as well as  the condition (\ref{cond}). These equations define the asymptotic density and pressure,    
\begin{equation}
\rho_a (t)=\lim_{r\to\infty}\rho(t,r)\,, \quad 	p_a(t)=\lim_{r\to\infty}p(t,r)\,,
\end{equation}
of the homogeneous perfect fluid of the asymptotic space-time ${\frak M}(a)$  we refer here as the FLRW fluid. The  natural splittings 
\begin{equation}
\rho(t,r)=\rho_a(t)+\delta\rho(t,r)\,, \quad p(t,r)=p_a(t)+\delta p(t,r)\,,	
\end{equation}
emphasizes the contribution of the point-dependent component of the perfect fluid of  ${\frak M}$ having the density $\delta\rho$ and pressure $ \delta p$.   

We remind the reader that the de Sitter expanding universe is the expanding portion ${\frak M}(a_{dS})$  of the de Sitter manifold having the de Sitter-Hubble constant frequency $\omega_{H}$ and the scale factor
\begin{equation}\label{aamk}
 a(t)\equiv a_{dS}(t)=e^{\omega_H(t-t_0)}   ~~~ \Rightarrow ~~~	\frac{\dot a(t)}{a(t)}=\omega_H\,,	
\end{equation}
which satisfies the initial condition $a_{dS}(t_0)=1$. The space of this manifold is infinite and expanding such that all the proper frames, including the black hole ones, have their own horizons on spheres of radius $r_a =\omega_H^{-1}$. These are event horizons separating different domain of causality. Particularly, the space-time ${\frak M}(a_{dS})$ is the asymptotic space-time of the Schwarzschild-de Sitter manifold with the Kottler metric (\ref{ss}) we denote from now by  ${\frak M}(a_{dS},m)$.

In the above geometries, the coordinate $t$ of the physical frames of the manifolds ${\frak M}$ is the cosmic time only in the physical domains where  $g_{00}(t,r)>0$. This means that the physical space domain is delimited by two dynamical spherical horizons, the black hole and cosmological ones whose radii, $r_b(t)$ and respectively $r_c(t)$,  have to be derived solving the equation $h(t,r)=1$ for expanding geometries and $h(t,r)=-1$ for collapsing ones \cite{Cot}. In many cases  these equations  have the desired positive solutions  only after a critical instant $t_{cr}$. Moreover, the space-time ${\frak M}(a)$ has its own apparent horizon on a sphere of radius 
\begin{equation}\label{ra}
	r_a(t)=\left| \frac{a(t)}{\dot a(t)}\right| 
\end{equation}
we call the asymptotic horizon. These horizons whose radii respect the hierarchy  $0<r_b(t)<r_c(t)<r_a(t)$ define a dynamical black hole  that can be observed only inside the physical domain, between the black hole and cosmological horizons.

\section{One-parameter models of dynamical black holes}

Let us consider now the new solutions of Einstein's equations with perfect fluid we proposed recently \cite{Cot1}. These are space-times which in physical proper frames have line elements of the form (\ref{fam}) defined by the functions, 
\begin{equation}\label{hdef}
	h_{\epsilon}(t,r)=-\frac{1}{3}\frac{\dot{M}(t)}{M(t)} r+\epsilon\sqrt{\frac{2 M(t)}{r}+\kappa^2 M(t)^2 r^2}\,,
\end{equation}
which depend on dynamical masses, $M(t)$,  its time derivative $\dot M(t)=\partial_t M(t)$,  and the  non-negative constant $\kappa=|\kappa|$  playing the role of free parameter. Here we introduce, in addition, the parameter $\epsilon=\pm 1$ for avoiding ambiguities when we extract the square root.  In Ref. \cite{Cot1} devoted to expanding geometries where $\epsilon=1$ this parameter was not considered explicitly.  We say  that  these geometries  are $\kappa$-models  denoting their space-times  by ${\frak M}(M,\kappa)$. Moreover, we have shown that these models  describe dynamical black hole having isotropic Einstein tensors whose components in proper physical frames have the form
\begin{eqnarray}
G^0_0(M,\kappa)&=&3\kappa^2 M(t)^2 +\frac{1}{3} \frac{\dot M(t)^2}{M(t)^2}-2\epsilon\dot{M}(t)\frac{1+{M(t)}\kappa^2r^{3} }{\sqrt{M(t)(M(t)\kappa^2r^3+2) r^3}}=8\pi \rho_{\kappa}\,,\label{E1}\\
G(M,\kappa)&=&3\kappa^2 M(t)^2+ \frac{\dot M(t)^2}{M(t)^2}-\frac{2}{3} \frac{\ddot M(t)}{M(t)}=-8\pi p_{\kappa}	\,,\label{E2}
\end{eqnarray}
satisfying the condition (\ref{cond}). We denoted here by $\rho_{\kappa}$ and $p_{\kappa}$ the density and pressure of the perfect fluids of our $\kappa$-models.  

For any such model the form of the function (\ref{hdef})  guarantees an asymptotic FLRW space-time ${\frak M}(a)$ whose Hubble function satisfies
\begin{equation}\label{at}
	\frac{\dot a(t)}{a(t)}=\lim_{r\to\infty}\frac{ h_{\epsilon}(t,r)}{r}=\epsilon\kappa M(t) -\frac{1}{3}\frac{\dot{M}(t)}{M(t)}\,.
\end{equation}
Integrating this equation  with the initial condition $a(t_0) =1$ we obtain the scale factor  
\begin{equation}\label{at1}
	a(t)=\left(  \frac{M_0}{M(t)}\right)^{\frac{1}{3}} \exp\left(\epsilon \kappa\int_{t_0}^t M(t')dt'\right)
\end{equation}
of the asymptotic FLRW space-time  where we denoted $M_0=M(t_0)$.
The Hubble functions must depend monotonously of time, without zeros in the physical time domain which  might produce singularities of the function $r_a(t)$ giving the radius of the asymptotic horizon (\ref{ra}).  We may prevent  these zeros to appear imposing different conditions  for expanding or collapsing  space-times as 
\begin{eqnarray}
	{\rm expanding:} &~~~~&\frac{\dot a(t)}{a(t)}>0 ~~\Rightarrow ~~ 	\frac{\dot{M}(t)}{M(t)}<0\,,\quad \epsilon=1\,,\label{expand}\\
	{\rm collapsing:} &~~~~&	\frac{\dot a(t)}{a(t)}<0 ~~\Rightarrow ~~ 	\frac{\dot{M}(t)}{M(t)}>0\,,\quad \epsilon=-1\,.\label{collaps}
\end{eqnarray} 
 In what follows we restrict ourselves to the expanding space-times which are of interest in cosmology. 

The physical meaning of the $\kappa$-models results from the right-handed sides of Einstein's equations (\ref{E1}) and (\ref{E2}). First of all, we observe that there are no static terms which means that we do not need to consider a cosmological constant, setting $\Lambda=0$. Furthermore, we separate the density and pressures of the FLRW fluid 
\begin{eqnarray}
\rho_a(t)&=&\frac{3}{8\pi}\left(\frac{\dot a}{a}\right)^2 =\frac{1}{8\pi}\left( 3\kappa^2 M(t)^2 +\frac{1}{3} \frac{\dot M(t)^2}{M(t)^2}-2\epsilon\kappa \dot M(t)\right)\,,\\
p_a(t)&=&- \frac{1}{8\pi}\left[ 3\left(\frac{\dot a}{a}\right)^2 +2\frac{d}{dt}\left(\frac{\dot a}{a}\right)   \right] =-\frac{1}{8\pi}\left( 3\kappa^2 M(t)^2+ \frac{\dot M(t)^2}{M(t)^2}-\frac{2}{3} \frac{\ddot M(t)}{M(t)} \right)\,,
\end{eqnarray}
which depend only on the Hubble function (\ref{at}). Hereby it results that the total density $\rho_{\kappa}=\rho_a + \delta\rho$  of the perfect fluid of the space-time ${\frak M}(M,\kappa)$ gets the new point-dependent  term 
\begin{equation}
	\delta\rho(t,r)=\frac{1}{4\pi}\epsilon \dot M(t)\left[  \kappa  -\frac{1+{M(t)}\kappa^2r^{3} }{\sqrt{M(t)(M(t)\kappa^2r^3+2) r^3}}   \right] \,, \label{dust}	
\end{equation}
while its pressure remains unchanged, $p_{\kappa}=p_a$. This means that $\delta\rho$ is the density of an amount of {\em dust} which does not modify te pressure $p_a$ of the FLRW fluid. 

We have thus the image of a system formed by a dynamical black hole of Schwarzschild type, surrounded by a cloud of dust hosted by the homogeneous perfect fluid of a FLRW space-time. It is remarkable that the features of all these components are determined only by the function $M(t)$ and the parameter $\kappa$. The black hole is produced by the typical  singular term  $\frac{2M(t)}{r}$ of the function (\ref{hdef}) while the dust density (\ref{dust}) was obtained solving the Einstein equations. Observing that this density also is singular in origin, behaving as $\sim r^{-\frac{2}{3}}$, we may ask which is the relation or  interaction between the black hole and dust predicted by our model.

A crucial step in solving this problem is to derive the total mass of dust, $\delta M(t)$, which, fortunately, is finite for $\kappa\not=0$ as the integral
\begin{eqnarray}
\delta M(t)=\int d^3 r\, \delta\rho(t, {\bf r})=4\pi \int_{0}^{\infty} dr\,r^2 \delta\rho(t,r) =
-\frac{1}{3\epsilon\kappa}\frac{\dot M(t)}{M(t)}\,,\label{dM}
\end{eqnarray}  
can be solved as in Appendix A. This remarkable result allows us to rewrite Eq. (\ref{at}) as
\begin{equation}\label{aMM}
\frac{\dot a(t)}{a(t)}=\epsilon\kappa\left[	M(t) +\delta M(t) \right]\,,
\end{equation}	
relating thus the principal quantities which determine the behaviour of our model. This relation can be seen as a conservation law or even as an equation of state we may use instead of the traditional one ($p=w\rho$) which does not make sense in the case of our $\kappa$-models where $\rho_{\kappa}$ is point-dependent while $p_{\kappa}$ is homogeneous.  Moreover, this result encourage us to define the total density of the system black hole-dust,  
\begin{equation}\label{totd}
	\hat\rho(t,{\bf r})=M(t)\delta^3({\bf r})+\delta\rho(t,r)\,,
\end{equation}
giving the total mass
\begin{equation}
\int d^3r\, 	\hat\rho(t,{\bf r})=M(t) +\delta M(t)	
\end{equation} 
of Eq. (\ref{aMM}). All these results hold only for $\kappa\not=0$ as for vanishing $\kappa$ the mass $\delta M$ diverges and  the above relations do not make sense.  The models with $\kappa=0$ which describe  Schwarzschild-type dynamical particles may be studied using specific methods  as in Ref.  \cite{Cot}.  

Finally, we must stress that the models which satisfy the conditions 
(\ref{expand}) and (\ref{collaps}) comply with the null energy condition, $\rho_{\kappa}+p_{\kappa}\ge 0$.  In our framework it is not difficult to find the remarkable identity 
\begin{equation}
\rho_{\kappa}(t,r)+p_{\kappa}(t)=-\frac{1}{4\pi r}\partial_t h_{\epsilon}(t,r)\,,
\end{equation}
which helps us to verify this property. 

\section{Black holes in de Sitter expanding universe}

Let us focus now on the models with $\epsilon=1$ in which the total mass $m$ of the system black hole-dust is conserved such that, according  to Eq. (\ref{aMM}), we may write the conservation law
\begin{equation}\label{mMM}
M(t)+\delta M(t)=M(t)-\frac{1}{3\kappa}\frac{\dot M(t)}{M(t)}=m \,.	
\end{equation}
Moreover, we see that the asymptotic FLRW space-time is just the expanding portion ${\frak M}(a_{dS})$  of the de Sitter manifold having the de Sitter-Hubble constant frequency $\omega_{H}=\kappa m$.  According to the standard interpretation, we assume that the FRLW fluid  is  {\em dark energy} of constant density and pressure that read 
\begin{equation}\label{dp}
\rho_{a_{dS}}(t)=-p_{a_{dS}}(t)=3\omega_{H}^2=3\kappa^2 m^2\,.	
\end{equation}
 The novelty here is that these quantities  are  produced by  the total mass $m$  and  our parameter $\kappa$ instead of a cosmological constant. We may say that this parameter  takes over the role of the cosmological constant assuring more flexibility to our approach where we do not need to consider global (or universal) quantities.  We have thus the advantage of constructing, for example, a de Sitter universe of Hubble constant $\omega_H$ that can be populated by a set of black holes  of parameters $(m_i,\kappa_i)$,  $i=1,2,...$,  if we take $\kappa_i=\omega_H m_i^{-1}$.

Furthermore, we focus on the dynamical components, the Schwarzschild-type black hole produced by the term $\frac{2 M(t)}{r}$ and the dust of density (\ref{dust}), assuming that  these are constituted by {\em ordinary matter}.  The function $M(t)$ which gives the dynamics of the model may be derived integrating the differential equation (\ref{mMM}) with the initial condition $M(t_0)=M_0$. Thus we obtain the black hole mass 
\begin{equation}\label{Mt}
\frac{M(t)}{M_0}=\frac{\mu}{1+e^{3\omega_H(t-t_0)}(\mu-1)}\,,\quad \mu=\frac{m}{M_0}\,,	
\end{equation} 
and the mass of the dust, $\frac{\delta M(t)}{M_0}=\mu-\frac{M(t)}{M_0}$, in units of $M_0$.	As we desire the functions $M(t)$ be smooth, we adopt the condition
\begin{equation}
\mu>1 ~~~ \Rightarrow~~~ \kappa\le \kappa_{lim}=\frac{\omega_H}{M_0}\,,	
\end{equation}
for avoiding time singularities during the black hole evolution.  We obtain thus two monotonous functions evolving between the limits
\begin{eqnarray}
\lim_{t\to -\infty}M(t)=m >&M(t)&> 	\lim_{t\to \infty}M(t)=0 \,,\\
\lim_{t\to -\infty}\delta M(t)=0 <&\delta M(t)&< 	\lim_{t\to \infty}\delta M(t)=m\,.
\end{eqnarray}
Plotting these functions as in Fig. 1 we obtain the image of the black hole evaporation when the black hole mass dissipates into dust such that the total mass $m$ of the system black hole-dust is conserved. Moreover, a rapid graphical analysis shows that the dust density is a non-negative function, $\delta\rho(t,r)\ge0$, with a singularity in $r=0$ which for $t\to \infty$ tends to a $\delta$-function that has to replace the initial singularity 
\begin{equation}
	\lim_{t\to-\infty}\hat\rho(t,{\bf r})=m \delta^3({\bf r})
\end{equation}
in the expression of the total density (\ref{totd}).   

 { \begin{figure}
		\centering
		\includegraphics[scale=0.35]{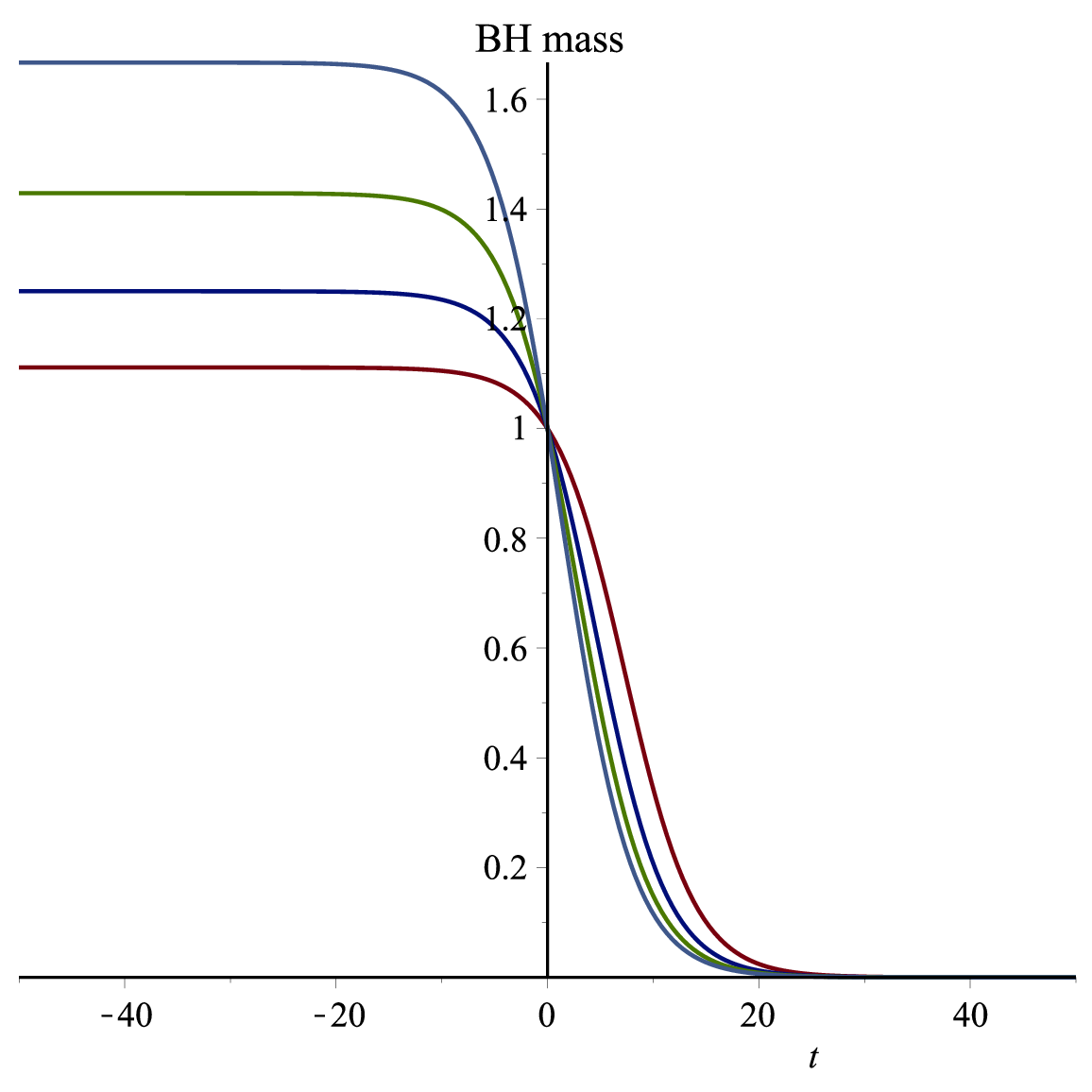}
		\includegraphics[scale=0.35]{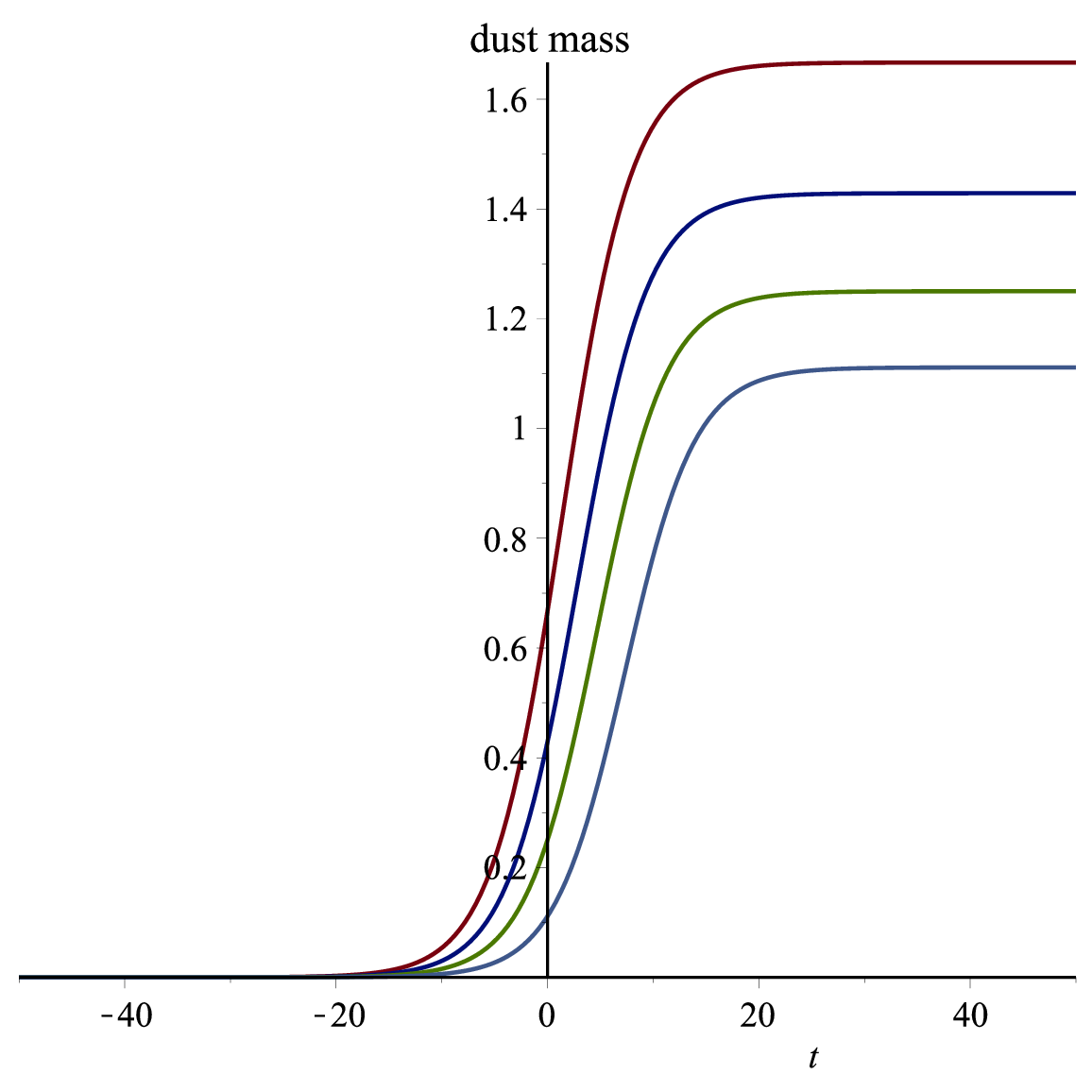}	
		\caption{The evolution of the functions $M(t)$  (left panel) and $\delta M(t)$ (right panel) in arbitrary units with $t_0=0$, $M_0=1$, $\omega_H=0.1$ and $m= 1.66, 1.42, 1.25, 1.11 < m_h=1.92 $, in descending order.}
\end{figure}}

The dynamics of the black hole model is encapsulated in the form of the function (\ref{hdef}) which  evolves between the limits
\begin{equation}
\lim_{t\to-\infty}h_{+}(t,r)=\sqrt{\frac{2m}{r}+\omega_H^2 r^2}\,, \quad
\lim_{t\to \infty}h_{+}(t,r)=\omega_H r\,. 	
\end{equation}
Therefore, the geometry evolves from an initial ({\em in}) state of a Schwarzschild-de Sitter black hole, concentrating the entire ordinary mass $m$, up to a final state ({\em out}) of an empty de Sitter expanding portion, proving thus  the complete black hole evaporation. It is known that in the $in$ state the black hole has the usual pair of  cosmological and black hole horizons only if $3\sqrt{3} m \omega_H<1$. As  Eq. (\ref{aamk}) allows us to substitute  $m=\omega_H\kappa^{-1}$ we obtain the  alternative conditions   
\begin{equation}\label{conk}
\kappa>\kappa_{h}=3\sqrt{3} \omega_H^2~~~~	{\rm or}~~~~m<m_h=\frac{1}{3\sqrt{3}\omega_H}\,,
\end{equation} 
necessary to meet in the $in$ state the entire system of  static horizons whose radii satisfy $0<r_{b|in}<r_{c|in}<r_a $.  For $0<\kappa <\kappa_h$ we remain only with the asymptotic event horizon of radius $r_a$ we also meet in the de Sitter $out$ state. Summarizing, we may say that the manifolds ${\frak M}(M,\kappa)$ evolve between the asymptotic limits, 
\begin{eqnarray}
{\frak M}_{|in}&=&\lim_{t\to-\infty}{\frak M}(M,\kappa)={\frak M}(a_{dS},m)\,,\\ 	
{\frak M}_{|out}&=&\lim_{t\to\infty}{\frak M}(M,\kappa)={\frak M}(a_{dS})\,,
\end{eqnarray}
during the evaporation of the mass $m$ into dust.  

{ \begin{figure}
		\centering
		\includegraphics[scale=0.35]{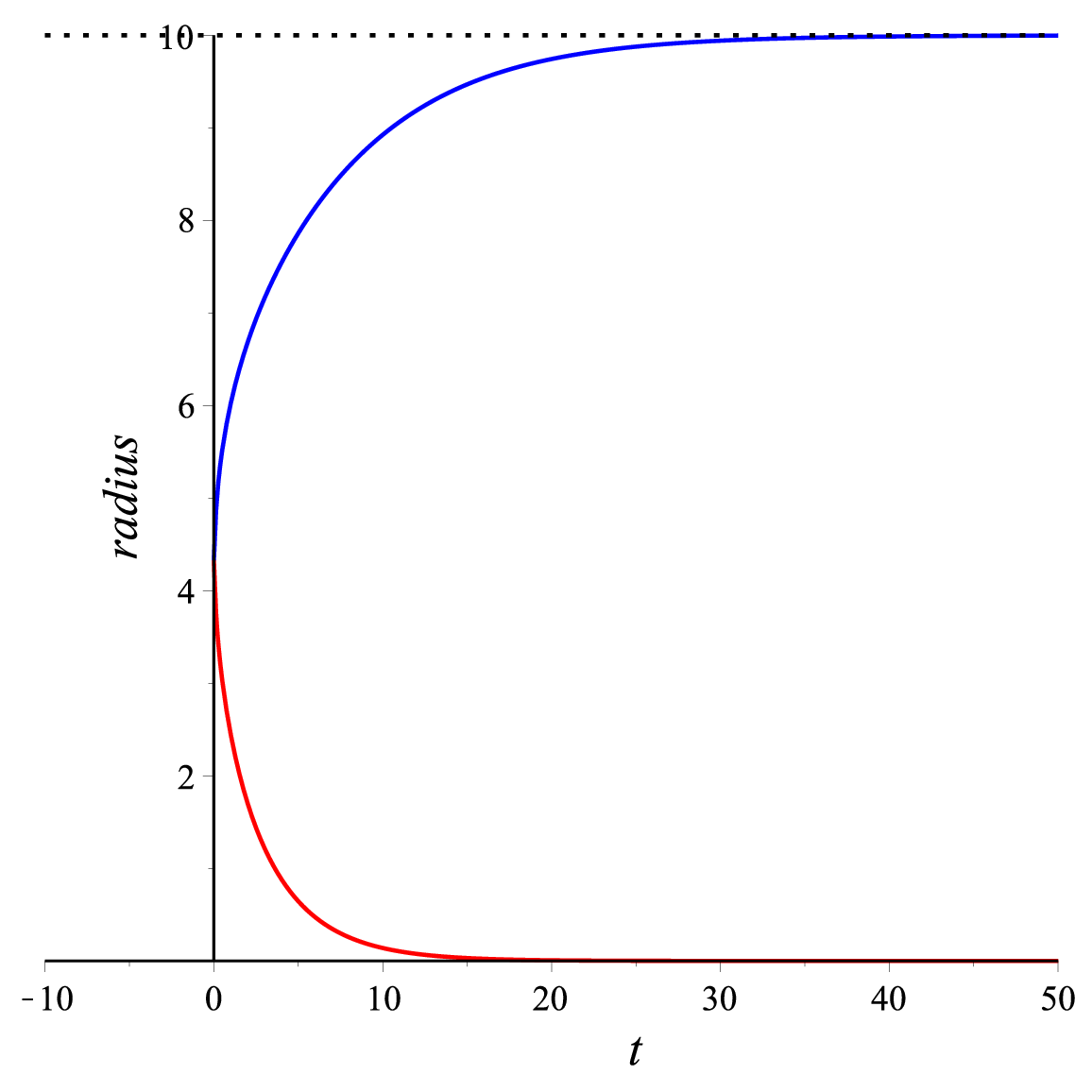}
		\includegraphics[scale=0.35]{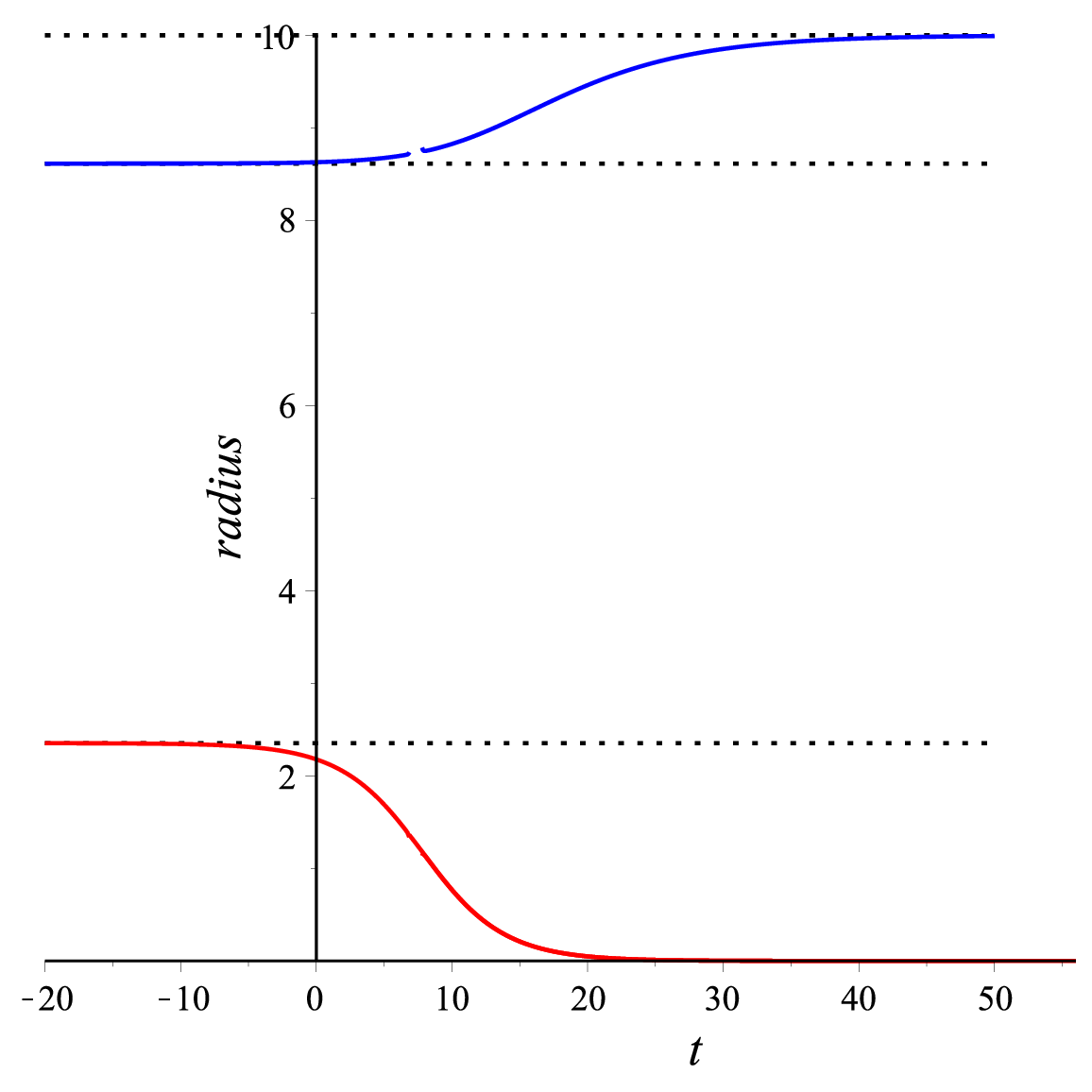}	
		\caption{The evolution of the cosmological (upper solid line) and black hole (lower solid line) horizons of a model with $t_0=0$, $M_0=1$, $\omega_H=0.1$, for which $\kappa_{lim}=0.1$ and $\kappa_h=0.0519$. For $\kappa=0.028$ the critical time is finite, $t_{cr}=t_0=0$, (left panel) while for $\kappa=0.09$ we have $t_{cr}\to-\infty$ and asymptotes  on the spheres of radii  $r_{b|jn} =2.35$ and $r_{c|in}=8.61$ (right panel). In both these cases the asymptotic horizon is a static sphere  having the radius $r_a=10$ (upper dotted line).}
\end{figure}}

At any finite instant $t$ the radii $r_b(t)$ and $r_c(t)$ of the dynamical horizons of  the black hole proper frame in ${\frak M}(M.\kappa)$ are solutions of the equation $h_+(t,r)=1$.  This equation can be solved as in the Appendix B obtaining solution with physical meaning only when the discriminant is positive, $\Delta\ge 0$. This happens  for $t \ge t_{cr}$ where $t_{cr}$ is the critical time that solves the equation $\Delta(t_{cr})=0$. The black hole and cosmological horizons arise at this time on the same sphere evolving then separately when the time is increasing. The cosmological horizon is approaching to the asymptotic one while the black hole horizon collapses to zero. These horizons are in fact event or trapping ones, evolving in accordance with the general black hole theory \cite{H1,H2} as the null energy condition is fulfilled,  
\begin{equation}
\rho_{\kappa}(t,r)+p_{\kappa}(t)=\rho_{a_{dS}}(t)+p_{a_{dS}}(t)+\delta\rho(t,r)\ge 0  \,.	
\end{equation}
because $\delta\rho\ge 0$ while the asymptotic density and pressure are given by Eq. (\ref{dp}).  
Under such circumstances, the black hole generates its own physical domain
\begin{eqnarray}
 {\frak P}=\{t,r, \theta, \phi\, |\,t>t_{cr}, r_b(t)<r<r_c(t)\} \subset{\frak M}(M,\kappa)	
\end{eqnarray}
in which $g_{00}(t,r)>0$ and $t$ is the cosmic time \cite{Cot1}.  Inside this domain an observer can measure the quantities indicating the presence of the dynamical black hole whose singularity in ${\bf r} =0$ remains hidden behind the black hole horizon.

The evolution of the black hole horizons  is strongly dependent on the value of  $\kappa\in(0,\kappa_{lim}]$. For $0<\kappa<\kappa_h$  we do not have horizons in the $in$ state such that $t_{cr}$ is finite and the radii of the dynamical horizons form usual C-curves as in the left panel of Fig. 2. However, for  $\kappa_h<\kappa<\kappa_{lim}$ the $in$ space-time has the complete system of horizons which are just the asymptotes at $t \to -\infty$ of the dynamical horizons which  evolve as in the right panel of Fig. 2. Only in this case we have the complete scenario of black hole evaporation from the $in$ to $out$ states, the dynamical masses behaving as in Fig. 1 where we respect the condition (\ref{conk}). Obviously, when $t_{cr}$ is finite, this scenario is cut, starting at this time with a black hole of mass $M(t_{cr})<m$ and an amount of dust of mass $\delta M(t_{cr})=m-M(t_{cr})$. 

We have thus the complete image of the time evolution of the dynamical black holes evaporating into dust, that can be observed  in  physical domains.  Outside these domains the system is a passive geometry without physical meaning, with $t$ playing the role of a  space-like coordinate,  but with mathematical properties that can be investigated. 

\section{Concluding remarks}

The general conclusion is that our $\kappa$- models with $\kappa\not=0$ describe systems formed by a Schwarzschild-type black hole  surrounded by a cloud of dust hosted by the perfect fluid of the asymptotic FLRW space-time. Any such system becomes active  at a critical instant $t_{cr}$ when the pair of  horizons appear on the same sphere,  evolving then in opposite directions,  creating thus the physical space between their spheres. 

Remarkably, these are genuine geometric models   without supplemental mechanisms of the  black hole thermodynamics. Their dynamics is determined only by the mass function and the parameter $\kappa$ which takes over the role of the cosmological constant. However, these parameters cannot be chosen  arbitrarily because we must prevent possible singularities   imposing supplemental conditions as in Eqs. (\ref{expand}) and (\ref{collaps}) which assure, in addition,  the null energy condition and a correct horizons dynamics. 

The problems that remain open are related to the quantities that can be measured by a remote observer co-moving with the FLRW fluid in the physical frame of the dynamical black hole. The only quantity that can be measured directly is the local dust density which is accessible in the physical domain. For emphasizing the dilation of the FLRW background and the presence of the black hole one may measure the redshift and the black hole shadow. We know the expressions of these quantities  at least in the $in$ state,  where the black hole is a Schwarzschild-de Sitter one \cite{Cot2}, or in the case of the  dynamical black holes with $\kappa=0$ we studied recently \cite{Cot}. However, for the models with $\kappa\not=0$ the algebraic calculations become extremely complicated such that the methods we applied in previous cases do not work.

Nevertheless, we hope to get over these difficulties  combining algebraic and numerical methods for  constructing realistic  models of expanding universe populated by evaporating black holes.

\appendix
\section{Integrating the dust density}
\setcounter{equation}{0} \renewcommand{\theequation}
{A.\arabic{equation}}

The integral of the dust density over a sphere of radius $R$ reads
\begin{eqnarray}
F(R)=4\pi \int_{0}^{R} dr\,r^2 \delta\rho(t,r)
 =\frac{1}{3}\epsilon\frac{\dot M(t)}{\sqrt{M(t)}} R\left(\kappa R^2\sqrt{M(t)}-\sqrt{R[M(t)\kappa^2 R^3+2]}     \right)\,.
\end{eqnarray} 
As the limit of this function  for $R\to \infty$ is indeterminate, we substitute $R\to\frac{1}{x}$ obtaining the limit 
\begin{equation}
\lim_{r\to\infty}F(R)=\lim_{x\to 0} F\left(\frac{1}{x} \right) =-\frac{1}{3\epsilon\kappa}\frac{\dot M(t)}{M(t)}\,,	
\end{equation}
we use in Eq. (\ref{dM}).

\section{Solving cubic equations}

\setcounter{equation}{0} \renewcommand{\theequation}
{B.\arabic{equation}}

For solving the cubic equation 
\begin{equation}
ar^3+br^2+cr+d=0\,, \label{e1}	
\end{equation}
 we substitute first  $r=x-\frac{b}{3a}$ obtaining the modified depressed equation $x^3-p x+q=0$ with the coefficients
\begin{eqnarray}
	p=\frac{b^2-3ac}{3a^2}\,,\quad q=\frac{27a^2d-9abc+2b^3}{27a^3}\,,
\end{eqnarray}
that can be solved using the following form of  Cardano's formulae 
\begin{eqnarray}
	x_1&=&\frac{1}{2}\left(A+\frac{4}{3}\frac{p}{A}\right)\,,\\
	x_2&=&\frac{1}{2}\left(Ae^{i\frac{2\pi}{3}}+\frac{4}{3}\frac{p}{A}e^{-i\frac{2\pi}{3}}\right)\,,\\
	x_3&=&\frac{1}{2}\left(Ae^{i\frac{4\pi}{3}}+\frac{4}{3}\frac{p}{A}e^{-i\frac{4\pi}{3}}\right)\,,	
\end{eqnarray}
where $	A=\left(\frac{2}{3}\right)^{\frac{2}{3}}\left[ i\sqrt{3}\sqrt{4p^3-27 q^2}-9 q \right]^{\frac{1}{3}}$. The cubic equations allow real solutions  only when their discriminants are positive, $\Delta=4 p^3-27 q^2 >0$. 

For our $\kappa$-models  the cubic equations $h_{\pm}(t,r)=\pm 1$ can be put in the canonical form (\ref{e1}) with the coefficients 
\begin{eqnarray}
&&a(t)=m\kappa^2[m-2M(t)]\,,\quad b(t)=2\epsilon\kappa[M(t)-m]\,,\\
&&c = 1\,, \hspace*{22mm} d(t)=-2M(t)\,.
\end{eqnarray}
 Finally the solutions we are looking for can be identified as
\begin{equation}\label{rbc}
r_b(t)=x_2(t)-\frac{b(t)}{3a(t)}\,,\quad 	r_c(t)=x_3(t)-\frac{b(t)}{3a(t)}\,,\quad \forall t\ge t_{cr}\,,
\end{equation}
while $x_1(t)<0$ is the nonphysical solution. Obviously, for $\kappa=0$ we recover the results of Ref. \cite{Cot}.

\end{document}